\documentclass[11pt]{article}

\usepackage[margin=1in]{geometry}
\usepackage{setspace}
\usepackage[T1]{fontenc}
\usepackage[utf8]{inputenc} 

\usepackage{graphicx}
\usepackage{booktabs}
\usepackage{array}
\usepackage{caption}
\usepackage{subcaption}

\usepackage{amsmath, amssymb}

\usepackage{siunitx}

\usepackage[hidelinks]{hyperref}

\usepackage[switch]{lineno}

\usepackage{authblk}

\usepackage[numbers,sort&compress]{natbib}

\usepackage{footmisc}
\AtBeginDocument{}

\usepackage{xcolor}
\definecolor{AnthropicOrange}{HTML}{CC785C} 

\definecolor{OpenAIColdGray}{HTML}{080808}
\definecolor{OpenAIWhite}{HTML}{FFFFFF} 

\definecolor{GrokCharcoal}{HTML}{313131}
\definecolor{GrokOffWhite}{HTML}{F8F7F6}    

\definecolor{PerplexityTurquoise}{HTML}{20808D} %

\definecolor{InsightTeal}{HTML}{2F7A73}
\definecolor{InsightTealTint}{HTML}{F3FAF9}

\definecolor{ContextSlateBlue}{HTML}{4C6A87}
\definecolor{ContextSlateBlueTint}{HTML}{F4F8FC}

\usepackage[most]{tcolorbox}

\usepackage{enumitem}

\setlist{
  topsep=3pt,        
  itemsep=1pt,       
  parsep=0pt,        
  partopsep=0pt
}

\usepackage{titlesec}

\titlespacing*{\section}
  {0pt}{0.9\baselineskip}{0.45\baselineskip}

\titlespacing*{\subsection}
  {0pt}{0.7\baselineskip}{0.35\baselineskip}

\titlespacing*{\subsubsection}
  {0pt}{0.55\baselineskip}{0.25\baselineskip}

\titleformat{\paragraph}[runin]
  {\normalfont\normalsize\bfseries}{\theparagraph}{0.75em}{}[.]

\titlespacing*{\paragraph}
  {0pt}      
  {0.25\baselineskip} 
  {0.5em}    

\usepackage{etoolbox}
\patchcmd{\thebibliography}{\setlength{\itemsep}{\z@}}{}{}{} 
\patchcmd{\thebibliography}{\setlength{\parsep}{\z@}}{}{}{}
\patchcmd{\thebibliography}{\setlength{\parskip}{\z@}}{}{}{}

\AtBeginEnvironment{thebibliography}{%
  \setlength{\itemsep}{0pt}%
  \setlength{\parsep}{0pt}%
  \setlength{\parskip}{0pt}%
}

\title{Can Consumer Chatbots Reason? A Student-Led Field Experiment Embedded in an ``AI-for-All'' Undergraduate Course}

\author[1,*]{Amarda Shehu}
\author[2]{Adonyas Ababu} 
\author[2]{Asma Akbary}
\author[2]{Griffin Allen} 
\author[2]{Aroush Baig} 
\author[2]{Tereana Battle}
\author[2]{Elias Beall} 
\author[2]{Christopher Byrom} 
\author[2]{Matt Dean} 
\author[2]{Kate Demarco} 
\author[2]{Ethan Douglass} 
\author[2]{Luis Granados} 
\author[2]{Layla Hantush} 
\author[2]{Andy Hay}
\author[2]{Eleanor Hay}
\author[2]{Caleb Jackson} 
\author[2]{Jaewon Jang}
\author[2]{Carter Jones} 
\author[2]{Quanyang Li}
\author[2]{Adrian Lopez}
\author[2]{Logan Massimo}
\author[2]{Garrett McMullin}
\author[2]{Ariana Mendoza Maldonado} 
\author[2]{Eman Mirza}
\author[2]{Hadiya Muddasar}
\author[2]{Sara Nuwayhid}
\author[2]{Brandon Pak}
\author[2]{Ashley Petty}
\author[2]{Dryden Rancourt}
\author[2]{Lily Rodriguez}
\author[2]{Corbin Rogers} 
\author[2]{Jacob Schiek}
\author[2]{Taeseo Seok}
\author[2]{Aarav Sethi}
\author[2]{Giovanni Vitela}
\author[2]{Winston Williams}
\author[2]{Jagan Yetukuri}

\affil[1]{UNIV 182 Course Designer and Instructor, Vice President and Chief AI Officer, George Mason University, Fairfax, Virginia, USA}
\affil[2]{Undergraduate Student, 
    George Mason University, Fairfax, Virginia, USA}
\affil[*]{Correspondence: \href{mailto:amarda@gmu.edu}{amarda@gmu.edu}}

\date{December 28, 2025} 

\begin{document}
\maketitle

\begin{abstract}
Claims about whether large language model (LLM) chatbots ``reason'' are typically debated using curated benchmarks and laboratory-style evaluation protocols. In this paper we report a complementary perspective: a student-led field experiment, embedded as a midterm project in UNIV~182 (AI4All) at George Mason University, a Mason Core course designed for undergraduates across disciplines with no expected prior STEM exposure.
Student teams designed their own reasoning tasks, ran them on widely-used consumer chatbots representative of current capabilities, and evaluated both (i) answer correctness and (ii) the validity of the chatbot's stated reasoning (e.g., cases where an answer is correct but the explanation is not, or vice versa).
Across eight teams that reported standardized scores, students contributed 80 original reasoning prompts spanning six categories (pattern completion, transformation rules, spatial/visual, quantitative, relational/logic, and analogical reasoning), producing 320 model responses plus follow-up explanations.
Aggregating team-level results, OpenAI GPT5 and Claude 4.5 had the highest mean answer accuracy (86.2\% and 83.8\%), followed by Grok 4 (82.5\%) and Perplexity (73.1\%); explanation validity showed similar ordering (81.2\%, 80.0\%, 77.5\%, 66.2\%).
Qualitatively, teams converged on a consistent error signature: strong performance on short, structured math/pattern items but reduced reliability on spatial/visual reasoning and multi-step transformations, with frequent ``sound right but reason wrong'' explanations. The assignment's primary contribution was pedagogical: it operationalized AI literacy as experimental practice (prompt design, measurement, rater disagreement, and interpretability/grounding), while producing a reusable, student-generated corpus of reasoning probes grounded in authentic end-user interaction.
\end{abstract}


\section{Introduction}
\label{sec:Introduction}

Whether contemporary large-language-model (LLM) chatbots ``reason'' remains contested. On the one hand, chain-of-thought prompting and related methods demonstrate that eliciting intermediate steps can improve performance on multi-step tasks and can produce outputs that \emph{look} like reasoning~\cite{wei2022cot}. On the other hand, benchmark design and measurement work cautions that high accuracy
on familiar task families may reflect learned priors rather than robust generalization~\cite{chollet2019measure}, and more recent analyses highlight ``collapse regimes'' in which reliability
degrades as problem complexity increases, even when models appear fluent and confident~\cite{shojaee2025illusion}.

This paper is motivated by a parallel question that arises \emph{outside} the research lab and crosses into the educational/classrooom experience: how can non-specialists evaluate chatbot reasoning claims in a disciplined manner, sot that they meaningfully understand the current state of reasoning? In practice, as an increasing number of studies report~\cite{}, undergraduates across all majors encounter consumer chatbots as productivity tools, study aids, and recommendation systems. Yet many students (and many instructors) lack a concrete evaluation framework for distinguishing (i) correct answers that are supported by coherent, constraint-following rationales from (ii) correct answers paired with confabulated explanations, or (iii) incorrect answers delivered with
high-confidence narratives. 

The need for such understanding is increasing. Adoption of these tools necessitate that we train students to be informed about growing capabilities (or lack of). In ongoing academic conversations, this need for understanding is subsumed under AI literacy initiatives. These initiatives, typically inclusive of undergraduates with no STEM backgrounds, have a grounding in general education. In this context, the broader question that AI literacy initiatives raise is pedagogical. The challenge is not only to introduce concepts, but to cultivate durable \emph{evaluation practice}: asking the right follow-up questions, documenting evidence, and making justified claims under uncertainty.

This challenge motivated the design of a new undergraduate course, a Mason Core course in IT \& Computing, UNIV 182, by the corresponding author of this paper. The course design and delivery will be documented elsewhere, but in this paper we hone in on one aspect that is representative of the course as a whole: the midterm project. The students that participated in this project agreed to co-authorship of this paper and to reporting for the broader community a classroom intervention designed to meet the AI literacy challenge. 

We note that UNIV 182 (AI4All) at George Mason University was designed as a Mason Core course for undergraduates across disciplines with no expected prior STEM exposure and was piloted in Fall 2025 with $40$ students. The experience in which we report in this paper was the midterm team-based project, which was framed as a field experiment on consumer chatbots. Teams designed their own reasoning tasks, executed a standardized protocol across multiple publicly available chatbots, and evaluated both (i) answer correctness and (ii) explanation validity (the logical soundness and constraint-consistency of the stated reasoning). The midterm was intentionally constructed
to do double duty: it functioned as an assessment \emph{and} (perhaps more importantly) as a guided learning experience in experimental
design, operationalization of constructs (``reasoning''), measurement, and interpretability/grounding. An important take-away was guided discovery of the current reasoning capabilities of state-of-the-art consumer chatbots by leading tech companies.

A key design choice was to treat students as \emph{investigators} rather than passive users. Instead of administering a fixed benchmark, the specification scaffolded prompt design through a taxonomy of reasoning categories (aligned to the Abstraction and Reasoning Corpus (ARC) perspective on generalization and priors), but granted students substantial freedom to craft tasks they believed would challenge consumer chatbots. This freedom produced a student-generated corpus with heterogeneity typical of real end-user interaction: prompts varied in creativity, ambiguity, and required inference. In a strict research context, that heterogeneity is a threat to validity for model ranking, but it is a crucial feature for pedagogy: students must confront construct validity (What are we \emph{actually} testing?), rater disagreement, and the distinction between ``sounding right'' and being right.

\paragraph{Contributions}
This is a non-typical paper that we believe makes the following contributions:
\begin{enumerate}
    \item A replicable assignment design for teaching reasoning evaluation in a general-education context,
    including a lightweight taxonomy of reasoning task types and a protocol for eliciting explanations.
    \item A student-generated corpus of reasoning prompts and annotated responses from consumer chatbots,
    created under a consistent classroom protocol.
    \item Descriptive empirical findings (not benchmark claims): cross-team patterns in answer accuracy,
    explanation validity, and recurring failure modes under end-user prompting.
    \item Pedagogical outcomes and implementation lessons for instructors building AI-literacy assessments
    that produce authentic artifacts and teach evaluation frameworks rather than tool familiarity.
\end{enumerate}

The rest of this paper is organized as follows. 
Section~\ref{sec:CourseContextAndLearningGoals} expands on the course context and learning outcomes that motivated the design of the midterm project. Section~\ref{Sec:MidtermProjectDesignAndSpecs} then provides details on the midterm specification, including the taxonomy of reasoning tasks and the documentation protocol. Section~\ref{sec:DataAndMethods} describes the artifact set analyzed and the evaluation criteria. Section~\ref{sec:Results} reports quantitative summaries and (importantly) qualitative vignettes that foreground student observations and the behaviors they found most instructive. Section~\ref{sec:PedagogicalOutcomes} synthesizes what students learned as evidenced in their artifacts and reflections.
Section~\ref{sec:Limitations} discusses limitations inherent to a classroom-based, student-led field evaluation of consumer chatbots.
Section~\ref{sec:Implications} draws broader implications for AI-literacy curriculum design and for assessments that treat evaluation practice as a core learning objective. Finally, Section~\ref{sec:Conclusion} concludes and outlines future directions, and Section~\ref{sec:Ethics} documents the consent, privacy, and data-availability considerations.
\section{Course Context and Learning Goals}
\label{sec:CourseContextAndLearningGoals}

UNIV 182 (AI4All) was designed as an ``AI-for-all'' undergraduate course for students across disciplines (first-year through senior), irrespective of prior STEM exposure. A central premise for its design was that AI literacy is not merely familiarity with tools, but a form of \emph{technical civic competence}: students should be able
to (i) understand how modern AI systems are built, (ii) what they can and cannot do, (iii) rigorously evaluate claims about system behavior, and (iv) responsibly use and critique AI-mediated information in academic and professional contexts. 

\subsection{In-class ``AI Studios'' as the Enabling Course Structure}

A distinguishing pedagogical feature of UNIV~182 was the routine use of extended, structured in-class work sessions that functioned as ``AI studios.'' Rather than allocating the full (twice weekly) $75$-minute class period to lecture only, several sessions were intentionally organized as active-learning blocks in which student teams worked on course projects in real time while the instructor circulated, provided feedback, and helped teams debug both technical and conceptual issues.
These studio sessions were used at key inflection points in the semester, including midterm project development and execution, final project scoping and prototyping, and preparation for structured debates.

The studios served three instructional purposes.

\begin{enumerate}

\item They created protected time for students from diverse majors and preparation levels to make progress on technical work without requiring extensive prior experience or out-of-class support networks.

\item They made methodological expectations visible and enforceable: teams could receive immediate feedback on whether a prompt was too familiar, whether an experimental comparison was controlled, whether a scoring rubric was defensible, and whether a claimed ``failure'' reflected model limitations or prompt ambiguity.

\item They reinforced the course’s ``AI literacy by doing'' premise by normalizing iterative practice (hypothesis, test, revision, and documentation) as the core learning loop.

\end{enumerate}

From the perspective of this paper, the studio structure is also important for replication. It provides a concrete implementation detail that helps explain why student-generated artifacts contain not only outputs, but also evidence of experimental reasoning, reconciliation of disagreements, and reflective interpretation.

\subsection{Mason Core Alignment: AI Literacy as General Education}

As a Mason Core course in Information Technology \& Computing, assessments were designed to satisfy at least one of the \emph{Learning Outcomes} (LO): (LO-1) understand principles of information storage, exchange, security, and privacy (and related ethical issues); (LO-2) consume digital information critically by selecting and evaluating relevant and trustworthy sources; (LO-3) use information and computing technologies to organize and analyze information and use it to guide decision-making; and (LO-4) choose and apply appropriate algorithmic methods to solve a problem~\cite{gmuMasonCoreIT}.

The midterm project was crafted to operationalize these outcomes in an authentic setting where students already had strong incentives to use chatbots. Rather than prohibiting consumer chatbots, the course treated
them as objects of inquiry. Students were asked to: design prompts (algorithmic thinking about rules and constraints), systematically apply identical prompts across tools (controlled comparison), record and organize evidence (documentation), evaluate explanation grounding (critical consumption), and reflect on risks of over-trust (ethics, security, and responsible use).

\subsection{Why a Chatbot Reasoning Midterm?}

Mid-semester is an ideal point to test whether students can synthesize: (i) conceptual knowledge of what LLM chatbots are trained to do, (ii) an understanding of reasoning as generalization under constraints (rather than mere fluency), and (iii) the practical skill of designing measurements. At this point in the course (due to intentional design), students have received technical understanding on how AI systems learn from data under different learning paradigms; how evaluation, bias, and failure modes arise; the foundations of deep learning architectures (including perceptrons, convolutional neural networks, and recurrent neural networks); and, critically, how self-attention and Transformer models underpin modern language models and chatbots, including the distinction between base language models and policy‑aligned conversational systems.

The students, therefore, were well positioned, and the midterm used the ``Can these chatbots reason?'' question as an integrative vehicle for teaching the following core components:
\begin{itemize}
    \item \textbf{Experimental design under constraints:} create a protocol, keep conditions comparable
    across models, anticipate confounds, and define success criteria.
    \item \textbf{Separation of outcome versus justification:} treat correctness and reasoning validity as
    distinct dimensions, because consumer chatbots can be correct for the wrong reasons (or vice versa). This distinction was inspired by recent work by Mitchell, albeit on the ARC benchmark~\cite{beger2025abstractreasoning}.
    \item \textbf{Critical AI literacy as practice:} verification behaviors, skepticism toward fluent narrative, and interpretation of results with appropriate caveats.
\end{itemize}

\subsection{Learning Objectives Targeted by the Midterm}
Within the Mason Core framework~\cite{gmuMasonCoreIT}, the midterm explicitly targeted three meta-skills:
\begin{enumerate}
    \item \textbf{Designing a measurable experiment:} operationalizing ``reasoning'' into observable proxies, applying consistent prompts across models, and documenting evidence. \\
    \emph{Mason Core alignment:} Outcome~(3) \textit{Use appropriate information and computing technologies to organize and analyze information and use it to guide decision-making.}

    \item \textbf{Evaluating grounding and constraint adherence:} judging whether explanations actually follow rules stated in the prompt and align with the produced answer. \\
    \emph{Mason Core alignment:} Outcome~(2) \textit{Consume digital information critically, capable of selecting and evaluating appropriate, relevant, and trustworthy sources of information.}

    \item \textbf{Developing durable evaluation habits:} recognizing when the appearance of reasoning is a function of prompt structure and when model behavior is brittle, inconsistent, or overconfident. \\
    \emph{Mason Core alignment:} Outcome~(2) \textit{Consume digital information critically, capable of selecting and evaluating appropriate, relevant, and trustworthy sources of information} (and, where risk/over-trust is discussed, Outcome~(1) \textit{Understand principles of security, privacy, and related ethical issues}).
\end{enumerate}

\subsection{Discovery-based Learning: Turning Evaluation into Lived Experience}

To keep these goals salient, the midterm specification framed the assignment as a (to-be-assessed) learning experience. The project was deliberately constructed as an exercise in \emph{discovery}: instead of having model limitations narrated to them in lecture(s), students were positioned to uncover those limitations through their own controlled interactions with consumer chatbots.

In practice, this meant that teams were expected to observe, firsthand, that seemingly minor changes in prompt wording, ordering, or constraint specification can materially alter performance; that some tasks are effectively ``too easy'' because they align with familiar training-set patterns or widely circulated examples; and that explanation fluency and confident tone are not reliability guarantees.
More broadly, the midterm was designed to shift ownership of learning from instructor to student.

By requiring students to generate their own probes, run comparisons, document evidence, and defend conclusions, the assignment created conditions under which students had to take learning agency into their own hands: they developed hypotheses, tested them, revised prompts, and reconciled disagreements based on artifacts rather than impressions.

This discovery-based structure is central to the course’s ``AI literacy by doing'' philosophy, because it turns abstract cautions about over-trust into concrete experiences that students can remember, reason about, and transfer to future use of AI systems.

\section{Midterm project design and specification}
\label{Sec:MidtermProjectDesignAndSpecs}

The midterm specification asked teams to compare four consumer chatbots—free versions (as of October 2025) offered by OpenAI, xAI (Grok), Anthropic (Claude), and Perplexity—and to identify ten reasoning tasks that would challenge models beyond familiar examples. Students were instructed to document each chatbot’s answer and, when an explanation
was not provided, to ask a follow-up prompt (e.g., ``How'' or ``Show me your reasoning'') to elicit the model’s rationale.

\subsection{Scaffolding Prompt Design: Taxonomy of Reasoning Tasks}

To support students who were new to experiment design, the specification provided categories of reasoning
tasks (and examples) aligned with the ARC perspective on generalization and priors \cite{chollet2019measure}. The specification explicitly noted that consumer chatbots are often trained on ``simple examples'' in each category and encouraged students to go beyond these familiar instances by creating prompts with non-obvious rules, multi-step transformations, or constraint interactions.

The six categories used for course scaffolding were:
\begin{enumerate}
    \item \textbf{Pattern completion:} identify and extend a pattern in sequences of numbers, letters, symbols,
    or structured objects. Harder instances often require recognizing a non-linear or nested pattern.
    \item \textbf{Transformation rules:} apply one or more deterministic rules to transform an input into an
    output (e.g., multi-step string edits or state transitions). These are particularly revealing when the
    prompt requires faithful execution of a procedure.
    \item \textbf{Spatial/visual reasoning:} reason about movement, rotation, reflection, or arrangement in
    space. Even when posed textually, these tasks probe mental simulation and constraint tracking.
    \item \textbf{Counting and quantitative reasoning:} compute totals, proportions, or invariants under rules
    (including embedded constraints). These can appear easy but become challenging with layered conditions.
    \item \textbf{Relational and logical reasoning:} infer consequences from partial orderings, implications,
    or relational statements (e.g., transitivity, syllogisms, and constraint satisfaction).
    \item \textbf{Analogical reasoning:} map relations across domains (A:B :: C:?), including cases where
    multiple plausible analogies exist and justification matters as much as the final mapping.
\end{enumerate}

Importantly, the category list was not meant as a rigid template; it was a shared vocabulary that allowed students to talk about ``types of reasoning'' and to notice systematic differences in model behavior across task families. Students were encouraged to be creative and go beyond these categories.

\subsection{Protocol: Controlled Comparison with Explanation Elicitation}

To keep the experiment interpretable at the classroom level, teams were also instructed to follow a common protocol:
\begin{itemize}
    \item Use the \textbf{same prompt} across all four chatbots.
    \item Record the \textbf{verbatim prompt}, the chatbot’s \textbf{final answer}, and the chatbot’s
    \textbf{explanation} (either provided initially or elicited via follow-up).
    \item Apply a team-defined scoring scheme for \textbf{answer correctness} and \textbf{explanation validity}.
\end{itemize}

This protocol was designed to encourage students to confront a central lesson: ``reasoning'' cannot be evaluated by surface fluency alone. When a model provides only an answer, an explanation follow-up often reveals whether the model is actually tracking constraints, or whether it is generating a plausible narrative after the fact.

\subsection{Team Structure as ``Force Multiplication''}

The midterm was team-based to make the evaluation effort feasible and to mirror real-world collaborative analysis. In many teams, members delegated responsibility for interacting with particular chatbots, enabling parallel data collection and comparative discussion. This structure also created opportunities for rater disagreement: teammates could debate whether an explanation truly followed the prompt’s intended semantics, and they could reconcile differences by returning to the task definition and constraints.

\subsection{Deliverables: Written Report and Class Presentation as Research-style Artifacts}

Teams submitted two deliverables:
\begin{enumerate}
    \item \textbf{A structured written report} documenting prompts and responses, with a summary of findings in the first two pages and task-by-task documentation thereafter.
    \item \textbf{A structured short presentation} (5 slides) summarizing the main findings as a meta-summary of the
    written report.
\end{enumerate}

This deliverable structure served two pedagogical goals. First, it required students to maintain an audit trail from claim $\rightarrow$ evidence $\rightarrow$ interpretation. Second, it trained students to
communicate technical findings concisely to peers, including stating caveats about ambiguity and prompt sensitivity.

\subsection{An Instructor-provided Example: From Toy Patterns to Adversarial Variants}

To illustrate the difference between toy problems and genuinely diagnostic probes, the specification provided examples of common prompt types that chatbots already handle well and noted that students should explore more interesting variants. The specification also included an example interaction with Grok involving a simple
transformation (``ABCD'' $\rightarrow$ ``ABCE'') and encouraged students to think along similar lines when designing their own prompts: small rule changes, multi-step transformations, and constraint interactions can elicit revealing failure modes.

\section{Data and methods}
\label{sec:DataAndMethods}

\subsection{Artifact set and Inclusion Criteria}

We analyze midterm artifacts (written reports and presentation summaries) produced by student teams as graded deliverables for the UNIV~182 midterm project (all contributing students consented to be co-authors on this manuscript). The artifact corpus includes student-authored reasoning prompts (documented in the Appendix), verbatim chatbot outputs (which we restrict as described in Section~\ref{sec:Ethics}), follow-up turns used to elicit explanations when a rationale was not initially provided, and team-authored evaluations and synthesis (shared in Section~\ref{sec:Results}).

We report two sets of results.

For \textbf{per-team reporting} and \textbf{qualitative analysis}, we include all teams with a written report and/or presentation that provides interpretable evidence of prompts and corresponding chatbot outputs.

For \textbf{quantitative aggregation}, we define a \emph{Quantitative Core (QC)} subset: teams that reported standardized numeric summaries (0--100\%) for both \textbf{answer correctness} and \textbf{explanation validity/grounding} as separate metrics for a comparable set of consumer chatbots.

Teams that did not report both metrics in a comparable form, or that used a non-standard model set (e.g., substituting Gemini for Claude), are included in per-team reporting and qualitative vignettes but excluded from QC mean calculations for the relevant chatbot/metric to avoid imputation or rubric retrofitting.

\subsection{Operationalizing ``Reasoning'': Correctness versus Explanation Validity}

Teams operationalized ``reasoning'' using two complementary, explicitly separated metrics:
\begin{enumerate}[leftmargin=*,itemsep=0.3em]
    \item \textbf{Answer correctness:} whether the final answer is correct under the prompt’s intended semantics.
    \item \textbf{Explanation validity (grounding):} whether the explanation is logically sound, adheres to stated constraints, and is consistent with the produced answer.
\end{enumerate}
This separation is essential in the classroom setting because it forces a distinction that novice users do not naturally make: a correct answer does not imply a correct rationale, and a coherent rationale can still contain a key inference error that leads to an incorrect answer.
Treating correctness and grounding as separate dimensions is therefore simultaneously a measurement decision and a learning objective.

\subsection{Scoring Practices and Rater effects}
Teams varied in scoring practice: some used multiple raters and reconciled disagreements through discussion; others used a single rater. We treat the resulting scores as student-annotated classroom measurements rather than ground-truth benchmark labels.
This framing is deliberate: the pedagogical goal is to teach students to argue from evidence, recognize ambiguity, and calibrate judgments.
Future iterations can strengthen reliability via shared calibration prompts and explicit inter-rater checks, but the current artifacts remain informative about how non-specialists evaluate chatbot reasoning in practice.

\subsection{Quantitative Aggregation and Qualitative Analysis}

For QC teams, we summarize performance by chatbot using an unweighted mean across eligible teams and report dispersion (standard deviation and range) to make variance visible. Because teams sometimes omitted a chatbot, substituted a different model, or reported only one metric, the effective sample size can differ by chatbot/metric; we therefore report $n$ explicitly in aggregate tables. All quantitative summaries are descriptive and should not be interpreted as definitive head-to-head benchmarks, since prompt sets differ across teams and model versions/settings are not controlled in consumer interfaces.

Complementing these summaries, we analyze student reports and slide decks qualitatively to identify (i) diagnostic failure modes, (ii) patterns of answer--explanation misalignment, (iii) prompt adaptations and follow-up strategies, and (iv) evidence of evaluation habits (verification, skepticism, and caveating).

Accordingly, Section~\ref{sec:Results} pairs aggregate summaries with qualitative vignettes that foreground students’ authentic observations and surprises, precisely the point where course concepts became concrete.
\section{Results}
\label{sec:Results}

We report two sets of results. First, we summarize quantitative patterns in answer accuracy and explanation validity across the eight teams that reported standardized metrics. Second (and more centrally for the goals of this paper), we foreground student work through qualitative vignettes drawn from written reports and slide decks, highlighting the specific behaviors that students found most instructive.

\subsection{Quantitative Summary and Variance Across Teams}

We begin with a descriptive quantitative summary over the \emph{Quantitative Core (QC)} subset defined in
Section~\ref{sec:DataAndMethods}. QC consists of teams that reported standardized numeric scores for both \textbf{answer correctness} and \textbf{explanation validity/grounding} as separate metrics, enabling
apples-to-apples aggregation without imputing missing values or retrofitting non-comparable rubrics.

\paragraph{Per-team results}
Table~\ref{tab:perteam} reports team-level outcomes for all graded submissions in our artifact corpus. Each entry is reported as \textbf{answer/explanation} percentage for the corresponding chatbot under a fixed set of ten prompts per team (ten prompts per team, one response per chatbot per prompt). ``--'' denotes that
a team did not evaluate that chatbot or did not report that metric. Teams with missing values are included for transparency, but are excluded from QC aggregation for the missing chatbot/metric. 

\begin{table}[htbp]
\centering
\caption{Per-team quantitative results from midterm submissions. Each entry is \textbf{answer/explanation} percentage. ``--'' denotes that the team did not evaluate that chatbot \emph{or} did not report that metric. Some submissions reported qualitative findings without standardized percentage scores; those are included here for completeness but excluded from quantitative aggregation in Table~\ref{tab:mean}. Some teams additionally crafted descriptive names for themselves.}
\label{tab:perteam}
\small
\setlength{\tabcolsep}{4.5pt}
\begin{tabular}{lccccc}
\toprule
\textbf{Team (submission)} & \textbf{OpenAI} & \textbf{Grok} & \textbf{Claude} & \textbf{Perplexity} & \textbf{Gemini} \\
\midrule
Team 2 & 100/95 & 100/90 & 60/60 & 90/100 & -- \\
Tool Tasks & 90/85 & 100/100 & 100/100 & 85/80 & -- \\
Team 5 & 100/60 & 90/70 & 80/40 & 70/30 & -- \\
Team DeMarco & 80/90 & 90/90 & 90/90 & 70/70 & -- \\
Carter Jones & 80/-- & 50/-- & 90/-- & 70/-- & -- \\
Reason Rangers & 90/90 & 50/50 & 80/80 & 60/60 & -- \\
Cookie Cutters & 70/70 & 80/80 & 80/90 & 80/70 & -- \\
Team One & 90/80 & 80/70 & -- & 80/70 & 90/90 \\
Error\ldots{}Data Not Found & 60/60 & 50/50 & 80/80 & 40/40 & -- \\
Byte Me & 100/100 & 100/90 & 100/100 & 90/80 & -- \\
EduVerse Squad & 90/70 & 100/60 & 100/80 & 100/80 & -- \\
\bottomrule
\end{tabular}
\end{table}

\begin{figure}[htbp]
\centering
\includegraphics[width=0.92\linewidth]{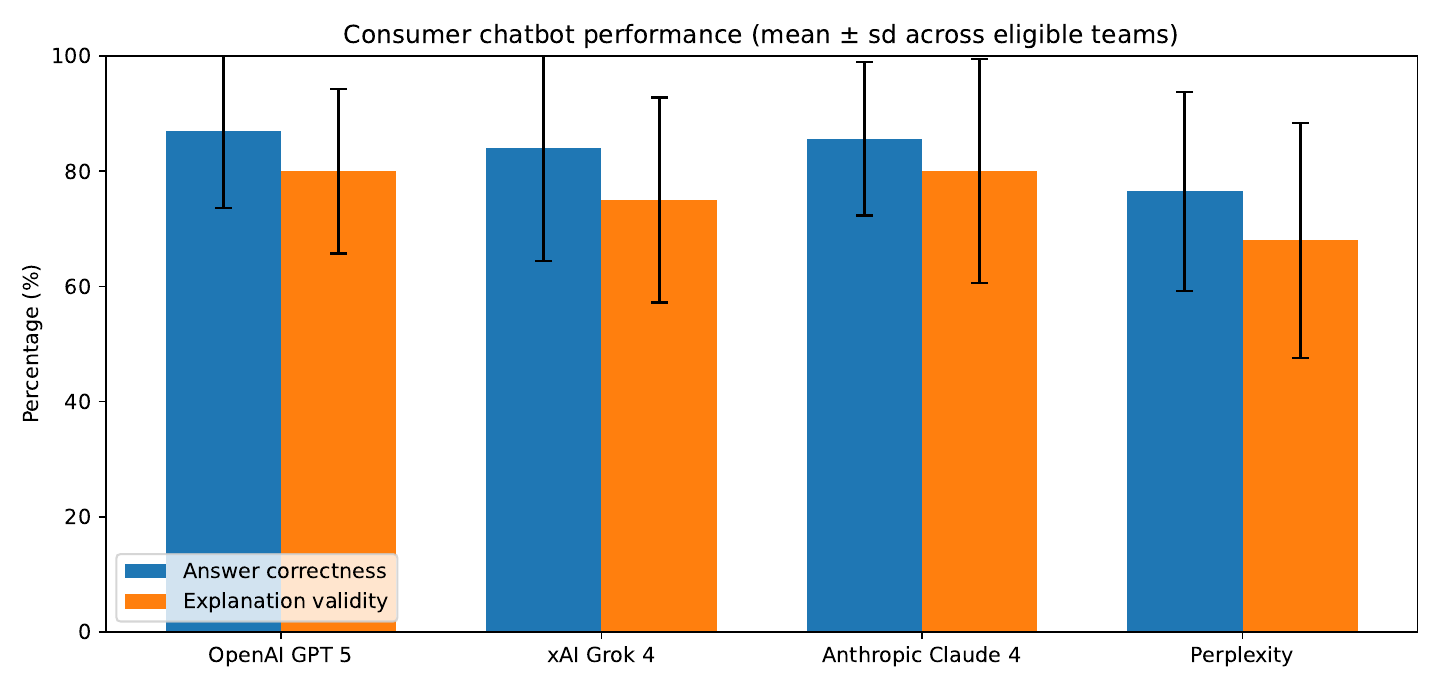}
\caption{Consumer chatbot performance aggregated across QC-eligible teams (mean $\pm$ sd): OpenAI GPT~5, xAI Grok~4, Anthropic Claude~4, and Perplexity.}
\label{fig:quant-summary}
\end{figure}

\paragraph{Aggregated results}
Table~\ref{tab:mean} now aggregates results by chatbot as an \emph{unweighted mean across eligible teams}, and
explicitly reports the number of contributing teams ($n$) for each chatbot/metric. This is a deliberate choice. Because some teams substituted an alternative chatbot (e.g., Gemini in place of Claude) or did not report explanation validity, the effective sample size differs across columns. We report dispersion (standard
deviation and range across teams) to make variance visible. Figure~\ref{fig:quant-summary} summarizes the tabular results visually.

\begin{table}[htbp]
\centering
\caption{Aggregate performance by chatbot over the QC-eligible subset for each chatbot/metric (unweighted mean across eligible teams). We report mean $\pm$ standard deviation and the observed range across teams. Values are percentages.}
\label{tab:mean}
\small
\setlength{\tabcolsep}{5.5pt}
\begin{tabular}{p{10em}p{4.5em}p{9em}p{9em}p{9em}}
\toprule

\textbf{Chatbot} & \textbf{$n$-teams} &
\textbf{Answer (\%)} &
\textbf{Explanation (\%) } &
\textbf{Gap (\%)} \\

& & \textbf{mean $\pm$ sd (range)} &
\textbf{mean $\pm$ sd (range)} &
\textbf{mean (range)} \\

\midrule
OpenAI GPT5 & 10 &
$87.0 \pm 13.4$ (60--100) &
$80.0 \pm 14.3$ (60--100) &
$+7.0$ ($-10$--40) \\
xAI Grok 4 & 10 &
$84.0 \pm 19.6$ (50--100) &
$75.0 \pm 17.8$ (50--100) &
$+9.0$ (0--40) \\
Anthropic Claude 4 & 9 &
$85.6 \pm 13.3$ (60--100) &
$80.0 \pm 19.4$ (40--100) &
$+5.6$ ($-10$--40) \\
Perplexity & 10 &
$76.5 \pm 17.3$ (40--100) &
$68.0 \pm 20.4$ (30--100) &
$+8.5$ ($-10$--40) \\
\bottomrule
\end{tabular}
\end{table}

Across the QC-eligible subset, OpenAI GPT5 and Anthropic Claude 4 exhibit the highest mean answer correctness and explanation
validity, followed by xAI Grok 4 and then Perplexity (Table~\ref{tab:mean}). However, variance across teams is
substantial. This variance reflects three interacting factors: (i) \textbf{prompt-set heterogeneity} (teams authored different reasoning tasks, with different degrees of ambiguity and difficulty), (ii) \textbf{prompt sensitivity} (small differences in wording, structure, or constraint emphasis can change behavior), and (iii) \textbf{rubric and rater effects} (teams applied slightly different thresholds when judging correctness versus grounding).

We emphasize that these quantitative summaries are \emph{descriptive} and should not be interpreted as a definitive benchmark ranking. The course context prioritizes authentic end-user interaction and
discovery-based evaluation over laboratory control: models were accessed through evolving consumer interfaces, tasks were student-authored, and scoring was performed by student teams. At the same time, the presence of substantial across-team variance is itself instructionally salient. It demonstrates to students (and to
readers) that claims about ``chatbot reasoning'' depend on experimental design choices and that disciplined evaluation requires controlling what can be controlled, documenting what cannot, and communicating caveats alongside conclusions, which we do next.

\subsection{Explanation Gaps are Common and Pedagogically Revealing}

A recurring pattern across teams is a measurable gap between answer correctness and explanation validity. Students repeatedly documented cases where a model reached the correct answer but justified it with invented rules, misread constraints, or post-hoc rationalization. These answer--explanation gaps matter for two reasons.

First, they expose a failure mode that casual users often miss: fluent explanations can create an illusion of grounded reasoning. Second, they provide a concrete mechanism for teaching interpretability and grounding in a general-education setting. Students learned to treat explanation validity as an important evaluation target.

One team explicitly summarized this phenomenon as cases where chatbots ``sound right'' but ``reason wrong,'' and treated such mismatches as evidence against human-like reasoning. Others observed the dual failure mode: explanations that correctly describe a rule but then execute it incorrectly in the final step (or vice versa).

\subsection{What Students Found Easiest: Short, Structured Arithmetic and Patterns}

Across teams, students reported strong performance across chatbots on short arithmetic, simple pattern completion, and tasks with explicit rules. This aligns with broader evidence that LLMs can perform well on structured math and pattern tasks with appropriate prompting~\cite{wei2022cot,cobbe2021gsm8k}. In the classroom context, this was an important baseline. Students learned to recognize when a prompt was likely too familiar and to revise prompts toward more diagnostic complexity.

Several teams also observed that strong performance in these domains could mask fragility: when a pattern was extended to include a second interacting rule, or when the task required maintaining state across multiple steps, accuracy dropped and explanations became inconsistent.

\subsection{What Students Found Hardest: Spatial/visual Reasoning and Multi-step Transformations}

Students across multiple teams independently designed tasks involving spatial movement, rotation/reflection, emoji patterns, grid navigation, or multi-step string transformations. These task types produced disproportionate failures and inconsistent reasoning narratives, consistent with research observations that increasing complexity and exact algorithmic requirements can trigger reliability collapse~\cite{shojaee2025illusion}.

This is also where student creativity was highest. When students moved beyond standard textbook patterns into ``adversarial-but-fair'' variants (e.g., multiple transformations, or spatial prompts that required careful indexing), they observed more frequent:
\begin{itemize}
    \item constraint violations (ignoring a stated rule),
    \item invented assumptions (adding rules not in the prompt),
    \item overconfident incorrect answers, and
    \item explanations that did not match the executed steps.
\end{itemize}

\subsection{Student Vignettes: What ``Failure'' Looked Like in Practice}

To make the above summative observations tangible, we present short vignettes that highlight the behaviors students found surprising, informative, or directly connected to course concepts about reasoning and grounding.

\begin{tcolorbox}[
  colback=AnthropicOrange!6!white,
  colframe=AnthropicOrange,
  title=\textbf{Claude’s ``thinking voice''},
  fonttitle=\bfseries]
\textbf{Fluent introspection without reliable convergence}

One team reported that Anthropic Claude 4 often began with high confidence and a clear ``I got this!'' posture, then
shifted into uncertainty (``I don’t understand.'') while producing long, human-like ``thinking'' text
(e.g., ``Let me try this---oh that’s not right.''). Students interpreted this as a form of simulated deliberation: the model can produce a convincing \emph{process narrative} even when it fails to converge on a correct solution, particularly on longer or noisier tasks.

\end{tcolorbox}

Pedagogically, this vignette was powerful because it externalized a key lesson: a model’s internal-looking monologue is not evidence of correct intermediate computation, and verbosity can be a (cognitively-taxing) liability when it substitutes for constraint tracking.

\begin{tcolorbox}[
  colback=GrokOffWhite,
  colframe=GrokCharcoal,
  title=\textbf{Grok’s iterative ``self-disagreement''},
  fonttitle=\bfseries]
\textbf{``Fighting with itself'' as an observable behavior}

Another team reported a striking Grok behavior in which the model questioned its own logic, declared its answer incorrect, and then repeated essentially the same reasoning process (multiple times) without resolving the underlying mistake. Students described this as the model ``fighting with itself'': producing self-correction signals without actually performing a corrective update to the procedure.

\end{tcolorbox}

This behavior became a concrete anchor for discussing the difference between \emph{meta-level} signals (``I might be wrong'') and \emph{object-level} correction (changing the steps that produced the error). It also helped students understand why ``reasoning layers'' can yield interesting observable behaviors without guaranteeing
robust, self-directed reasoning.

\begin{tcolorbox}[
  colback=OpenAIWhite,
  colframe=OpenAIColdGray,
  title=\textbf{Prompt length and overcomplication},
  fonttitle=\bfseries]
\textbf{When more context actually reduced accuracy}
The Tool Tasks team reported that as more prompts (and hence more total interaction context) were introduced, accuracy and reasoning quality decreased, particularly for ChatGPT and Perplexity. They also reported that an emoji-based spatial/visual reasoning task ``tripped up'' roughly half the tools they tested.
\end{tcolorbox}

This vignette reinforces two practical lessons for AI literacy. First, more text is not always more signal: additional context can introduce opportunities for misinterpretation or spurious rule invention. Second,
emoji/spatial prompts provide a low-barrier way for students to probe brittle symbolic and positional reasoning without requiring specialized mathematics.

\begin{tcolorbox}[
  colback=InsightTealTint,
  colframe=InsightTeal,
  title=\textbf{Rules versus generalization},
  fonttitle=\bfseries]

\textbf{Pattern followers, not reliable general thinkers}

A different team summarized a class-level takeaway succinctly: the models they tested were ``good pattern followers, not reliable general thinkers,'' and users ``can’t trust confidence or nice wording as proof the reasoning is correct''; verification remains necessary.
\end{tcolorbox}

This theme appeared in multiple forms across teams:
\begin{itemize}
    \item models that correctly restated a rule but executed it incorrectly on a second variant,
    \item models that succeeded on a spatial rule in one representation but failed when the same rule was
    expressed with emojis or a slightly different format, and
    \item models that produced a correct answer via a shortcut but could not justify it under the prompt’s
    intended semantics.
\end{itemize}

For the course, this is the point where ``AI literacy by doing'' became visible: students moved from evaluating outputs as consumers to evaluating systems as investigators, with a bias toward reproducible evidence.

\begin{tcolorbox}[
  colback=ContextSlateBlueTint,
  colframe=ContextSlateBlue,
  title=\textbf{Context richness as a mediator of ``apparent reasoning''},
  fonttitle=\bfseries]
The EduVerse team explicitly concluded that chatbot reasoning appeared highly dependent on the ``clarity and depth of context'' provided in the prompt: when context was rich and structured, models produced more accurate and coherent explanations; when prompts were vague, models drifted into assumptions or over-analysis.
\end{tcolorbox}

This vignette matters because it connects student experience directly to a broader interpretability claim: what end-users perceive as ``reasoning'' is often a co-production between human prompt structure and model pattern completion. In the classroom, this became a constructive outcome rather than a critique: students learned
to treat prompt engineering as experimental control and to recognize when a result is driven by framing.

\subsection{Student Conclusions: Can Consumer Chatbots Reason?}

Across teams, conclusions were nuanced rather than binary. Many teams converged on a conditional claim: consumer chatbots can often perform well on short, structured tasks with explicit rules, but they are brittle on tasks requiring spatial simulation, multi-step transformations, strict constraint tracking, or generalization
under format shifts. Several teams therefore characterized chatbot behavior as \emph{imitation} of reasoning: plausible narratives plus partial rule-following, without consistent reliability across diverse problem types.

Importantly, this nuance reflects learning. Students did not merely rank tools; they articulated what they mean by ``reasoning,'' defended operational definitions using evidence, and properly caveated their claims based on prompt ambiguity, scoring subjectivity, and the limitations of consumer-facing interfaces. In that sense, the most significant result of the midterm project was not a leaderboard, but the emergence of evaluative practice: students learned how to ask disciplined questions of AI systems and how to communicate evidence-based interpretations.

\section{Pedagogical Outcomes: AI Literacy as Experimental Practice}
\label{sec:PedagogicalOutcomes}

A central motivation of UNIV~182 (AI4All) is to move beyond \emph{tool familiarity} (e.g., knowing which chatbot exists) toward \emph{evaluative competence}: the ability to form justified beliefs about what a model can and cannot do, under what conditions, and with what risks. The midterm project operationalized this goal by positioning consumer chatbots as empirical objects of study.
Students were not asked to \emph{use} chatbots to complete a task; they were asked to \emph{test} chatbots, build a measurement protocol, and defend an evidence-based claim.

This section synthesizes pedagogical outcomes that are visible in the submitted artifacts (reports and presentations) and in the work practices the assignment required (prompt design, controlled comparisons, measurement, reconciliation of disagreements, and communication of results).

\subsection{From ``AI as magic'' to ``AI as an Artifact You Can Test''}

Many undergraduates encounter AI systems first as high-agency assistants that produce fluent text on demand. A recurring instructional challenge in general-education contexts is that fluency can be misread as correctness and, more subtly, as \emph{understanding}. The midterm intervenes by forcing a shift in stance: students treat the chatbot output as a \emph{claim} that needs evaluation, not an \emph{authority} that settles the matter.

Concretely, the project contains three deliberate design components that push students into an experimental mindset:

\begin{enumerate}[leftmargin=*,itemsep=0.25em]
\item \textbf{Students author the probes.} Instead of relying on instructor-provided prompts, students design tasks that they believe will stress-test models, which quickly reveals that ``reasoning'' is not a single capability but a family of behaviors that can be elicited (or masked) by prompt form.
\item \textbf{Students standardize inputs across systems.} Each prompt is executed on multiple chatbots under a shared protocol, making differences in output salient and discouraging ``single-example'' conclusions.
\item \textbf{Students evaluate both outputs and explanations.} Students explicitly separate \emph{answer correctness} from \emph{explanation validity}, creating space to notice the signature failure mode of consumer systems: outputs that are persuasive yet poorly grounded.
\end{enumerate}

\subsection{Mapping to Mason Core IT \& Computing Learning Outcomes}

Because UNIV~182 is a Mason Core course, the midterm project was deliberately designed to satisfy multiple IT \& Computing learning outcomes \cite{gmuMasonCoreIT}. Table~\ref{tab:masoncore-mapping} provides a concrete mapping between (i) the observable activities demanded by the project and (ii) the corresponding Mason Core outcomes.

\begin{table}[t]
\centering
\caption{How the midterm project operationalizes Mason Core IT \& Computing outcomes \cite{gmuMasonCoreIT}.}
\label{tab:masoncore-mapping}
\small
\begin{tabular}{p{2.2cm}p{5.0cm}p{6.0cm}}
\toprule
\textbf{Outcome} & \textbf{Midterm activity} & \textbf{Observable evidence in student artifacts} \\
\midrule
(1) Security, privacy, ethics & Discussing responsible use, uncertainty, and when to trust/verify & Teams explicitly warn against over-trusting fluent outputs and note ethical risks of dependence and overconfidence. \\
\addlinespace
(2) Critical consumption of digital information & Treating chatbot outputs as claims; checking for contradictions, hidden assumptions, and unsupported steps & Students identify ``sound right but reason wrong'' explanations and document disagreement between answers and rationales. \\
\addlinespace
(3) Organize/analyze information for decision-making & Logging outputs, computing accuracy/validity metrics, summarizing patterns across tools & Teams produce tables with accuracy and explanation-validity rates and use them to justify comparative claims. \\
\addlinespace
(4) Apply algorithmic methods & Designing tasks from reasoning categories; specifying rules; analyzing rule-following vs generalization & Students create prompts that require multi-step transformations, constrained logic, and compositional patterns and then diagnose where rule discipline breaks. \\
\bottomrule
\end{tabular}
\end{table}

\subsection{What Students Learned about ``Reasoning'' by Having to Measure It}

A distinctive feature of the project is that it forces students to transform a vague question (``Can chatbots reason?'') into an operational definition with measurable proxies. Across teams, three measurement lessons recur.

\paragraph{(1) Reasoning is Plural, not Singular}
The category scaffold (pattern completion; transformation rules; spatial/visual; quantitative; relational/logic; analogical) functioned as a practical taxonomy. Students quickly observed that systems can appear strong in one category and brittle in another, which is pedagogically valuable because it counters both naive optimism (``it can do everything'') and naive dismissal (``it can do nothing'').

\paragraph{(2) Correct answers can be achieved for the wrong reasons}
Separating \emph{answer correctness} from \emph{explanation validity} required students to define what counts as a valid justification: rule compliance, internal consistency, and alignment between intermediate steps and the final claim. This separation is not merely a grading artifact; it is a transferable AI literacy skill, because many real-world harms arise not from a wrong final answer per se but from a wrong justification that is trusted and reused.

\paragraph{(3) Prompt design is part of the phenomenon, not a nuisance variable} Students reported that minor changes in wording, specificity, or constraints could change both answers and explanations.
Rather than treating this as a failure of the evaluation, the assignment treats it as the point: in real-world use, \emph{end users control prompts}, and thus prompt sensitivity is part of the system's practical reliability profile.

\subsection{Failure Modes as Teachable Moments}

Beyond category-level performance differences, students encountered behaviors that made the class content feel concrete.

\paragraph{Insistence, Overconfidence, and ``Interactional'' Failure}

Several teams reported cases where a chatbot confidently defended an incorrect solution path, even when challenged, and produced long explanations that did not resolve the contradiction. In classroom discussion and in some written artifacts, students described these dynamics using everyday social language (e.g., ``arguing,'' ``doubling down,'' or framing the interaction as the model ``fighting with itself''). Pedagogically, this mattered, because students saw that a fluent explanation is not a guarantee of epistemic humility, calibration, or truth-tracking.

\paragraph{Over-reasoning and invented rules} A common student diagnosis, especially on tasks with crisp constraints, was that some systems introduced extra assumptions or additional mechanisms not present in the prompt. Students learned to treat this not as ``creativity'' but as a form of constraint violation, which is precisely the distinction that matters when using AI systems for decision support.

\paragraph{Sycophancy and social steering} At least one team explicitly labeled a model's tendency to shift answers in response to user pressure as ``sycophancy,'' using it as evidence against robust reasoning. This is an important literacy outcome: it reframes chatbot helpfulness as a potential failure mode when truth and justification matter.

\subsection{Team Process}

The project design intentionally leveraged teamwork to approximate a small-scale research workflow. Teams delegated chatbots to different members, ran parallel trials, and then reconciled results in shared documents. This workflow produced a concrete lesson about reproducibility: even when prompts are identical, differences in interface defaults, follow-up prompting, and interpretation can yield divergent outcomes. Students experienced, in miniature, why scientific claims about AI need protocols and why evaluation is not merely ``asking questions'' but \emph{controlling interaction}.

\subsection{Communication Outcomes: Making Technical Claims Accessible}

Finally, the five-slide presentation constraint required students to distill technical work into a structured argument: what was tested, what was observed, what evidence supports the claim, and what caveats limit generalization. For a general-education course, this is a non-trivial achievement: students practiced translating a messy interactional phenomenon into falsifiable claims with quantitative summaries and qualitative examples.

In summary, the midterm project functioned as ``learning disguised as assessment'' by requiring students to \emph{do} evaluation rather than merely \emph{learn about} evaluation. The outputs are not only grades but rather artifacts of a repeatable classroom research practice.

\section{Limitations}
\label{sec:Limitations}

This paper intentionally reports a \emph{field experiment embedded in a course}, not a controlled benchmark study. The resulting evidence is authentic to end-user interaction, but it inherits limitations that matter for interpretation. We group these as standard threats to validity (construct, internal, external, and conclusion validity), plus constraints imposed by consumer chatbot interfaces.

\subsection{What this Study is and is not}

\paragraph{Not a head-to-head benchmark}
Teams authored different prompts, used slightly different rubrics, and interacted with chatbots through free consumer interfaces that may vary in defaults (e.g., memory, tool use, and response style).
Accordingly, aggregate performance numbers should be read as descriptive summaries of student-generated evaluations, not as definitive rankings.

\paragraph{A pedagogical intervention with empirical byproducts}
The primary aim is educational: to teach students to evaluate AI reasoning claims with discipline. Empirical results are a meaningful byproduct, but the design prioritizes learning objectives over experimental control.

\subsection{Construct Validity: What do the Prompts Actually Measure?}

\paragraph{Reasoning versus world knowledge versus tool behavior}
Some prompts may implicitly test retrieval (e.g., facts or geography), interface affordances (e.g., whether a system browses), or conversational strategies (e.g., hedging, deflection), not reasoning in a narrow cognitive sense. This is not necessarily a flaw: consumer chatbots are deployed as composite systems. However, it complicates any claim that the project isolates ``reasoning'' as a single latent ability.

\paragraph{Ambiguity and multiple valid interpretations}
Some reasoning prompts admit multiple plausible solutions (especially analogies, creative transformations, and ill-posed spatial descriptions). Student scoring protocols varied in how they handled alternative but defensible answers. This affects both correctness estimates and explanation-validity estimates.

\paragraph{``Explanation validity'' is an operational proxy}
A key methodological choice is to evaluate the \emph{stated} explanation, not an internal chain-of-thought. Explanations in consumer chatbots may be post-hoc rationalizations, abbreviated summaries, or policy-filtered outputs. Thus, explanation validity measures the quality of the \emph{justification presented to users}, which is pedagogically and practically important, but it is not a direct window into internal computation.

\subsection{Internal Validity: Sources of Variation in Interaction}

\paragraph{Model versions and settings were not controlled}
Students used free, publicly available consumer interfaces during a specific window in the semester. Model versions may have changed across days or even within the same day. Some interfaces may enable tools, memory, or personalization by default, which can alter outputs.

\paragraph{Follow-up prompting and ``help'' effects}
The protocol asked students to request reasoning if not provided; in practice, follow-up prompts sometimes included hints, clarifications, or corrections. This interaction is realistic (users do provide feedback) but it introduces dependence between the initial response and later responses and so makes it difficult to interpret a single score as a one-shot capability measure.

\paragraph{Non-independence and learning effects within sessions}
When multiple prompts are asked in sequence, chatbots may implicitly condition on earlier context. Even when students attempted to reset context, differences in session handling could persist.

\subsection{External Validity: Generalizability Beyond this Course}

\paragraph{Single institution and a specific student population}
Results reflect one course at one institution in one semester, with the particular distribution of majors, years, and student interests present in UNIV~182. Replication across institutions and semesters is needed to assess the generality of both (i) the pedagogical outcomes and (ii) the observed chatbot failure modes. We note, though, that generality was not the objective of this paper. Rather, the objective was to report on an interesting AI literacy experiment that engages undergraduate students and turns the process of assessment in team-based learning and discovery.

\paragraph{Prompt set is shaped by student creativity}
A defining feature of the intervention is that students deliberately sought ``interesting'' failure cases. This yields valuable probes for model brittleness, but it also means the prompt distribution is not representative of typical consumer usage.

\subsection{Conclusion validity: scoring reliability and aggregation}

\paragraph{Rater subjectivity and differing rubrics} 
Some teams used multiple raters and reconciled disagreements; others used a single rater. Even when teams used similar concepts (correctness and explanation validity), thresholds differed. As a result, unweighted averaging across teams should be treated as an illustrative summary rather than a statistically rigorous estimate.

\paragraph{Small sample size at the team level}
With ten prompts per team, a single ambiguous or disputed item can change percentages materially. This is not a defect in the classroom context, but it limits any inferential claims.

\subsection{Constraints and Confounds from Consumer Chatbot Interfaces}

\paragraph{Interface-Mediated Behavior is Part of the System}
Perplexity, for example, may present citations and external context; other systems may prioritize conversational tone or structured reasoning templates. These interface-layer choices influence how ``reasoning'' appears to users and therefore influence both student judgments and real-world trust calibration.

\paragraph{Terms of Service and Observability Limitations}
Consumer platforms may restrict data collection, rate limits, or visibility into system components. These constraints limit the completeness of logging and the feasibility of fully reproducible protocols.

\paragraph{Summary}
Taken together, these limitations suggest two practical takeaways for instructors who may wish to replicate or adapt this midterm experience: (1) interpret quantitative summaries as descriptive indicators that are likely to vary with cohort, tools, and local course context rather than as generalizable effect estimates, and (2) plan to capture and report qualitative evidence, such as failure modes, prompting trajectories, and student reflections, as core outcomes that make the exercise instructional and portable.
\section{Implications for Course and Assessment Design}
\label{sec:Implications}

The intervention described here sits at the intersection of three communities that often speak past one another: (i) AI researchers debating ``reasoning'' in models, (ii) educators attempting to teach AI literacy at scale, and (iii) everyday users encountering consumer chatbots as general-purpose assistants. We outline implications for each, emphasizing the paper's dual identity as a pedagogical report and a field-style evaluation of deployed systems.

\subsection{Implications for AI Literacy in General Education}

\paragraph{AI literacy should include evaluation practice, not just concepts}
Students can memorize that LLMs are probabilistic token predictors and still over-trust fluent explanations. The midterm shows that evaluation habits (verification, skepticism, and rubric-based judgment) can be taught through structured practice even in a mixed-background classroom.

\paragraph{A category scaffold is a practical bridge into technical ideas}
Reasoning categories provided a lightweight, non-intimidating entry point into a technical notion: different task families place different demands on representation, compositionality, and constraint satisfaction. Students who entered the course without STEM backgrounds (which was the overwhelming majority of the students in this course) were still able to design sophisticated stress tests when given a taxonomy and examples.

\paragraph{Separating ``answer'' from ``reason'' is a durable literacy habit}
Consumer chatbots increasingly present explanations by default. If students learn only to check answers, they miss the more subtle risk: explanations that are rhetorically persuasive but logically invalid.
The project foregrounds this separation as a core skill for responsible use. It is worth noting that this is a durable skill, increasingly an important conversation in academic settings on what to teach students as AI technologies evolve apace.

\subsection{Implications for Assessment Design in the Era of Ubiquitous Chatbots}

\paragraph{Turning chatbots into the object of analysis reduces incentives for misuse}
Traditional assessments can be undermined by outsourcing work to chatbots. In contrast, this midterm makes chatbot interaction the required substrate. Students gain experience with the tool while practicing critical distance, and academic integrity concerns shift from ``did you use AI?'' to ``did you evaluate it responsibly and report accurately?''

\paragraph{Assessment-as-inquiry scales beyond majors}
Because the tasks are authored by students and grounded in consumer tools they already encounter, the intervention can scale to non-majors without requiring advanced prerequisites. This suggests a general strategy for Mason Core-style (and, more broadly, general education) courses: use inquiry-based assignments where the technical object is observable and testable.

\paragraph{Rubrics and calibration examples matter}
One lesson from the heterogeneous scoring approaches is that shared calibration improves reliability and also improves learning: students become more articulate about what counts as a valid justification, what counts as ambiguity, and what constitutes a constraint violation.

\subsection{Implications for Evaluating Consumer Chatbots}

\paragraph{Field-style evaluation complements benchmarks}
Benchmarks are essential for controlled comparisons, but they often abstract away the interactional features that dominate real-world use: follow-up prompting, ambiguity negotiation, refusal behavior, and persistence in incorrect answers. Student-designed prompts and dialogues capture aspects of deployed behavior that benchmarks typically miss.

\paragraph{Explanations are part of the product surface and must be evaluated as such}
In consumer settings, explanations are not merely interpretability artifacts; they are persuasion mechanisms. A system that produces correct answers with systematically unreliable explanations can be more harmful than a system that admits uncertainty, because it creates false confidence.

\paragraph{Prompt sensitivity should be treated as a reliability dimension}
Students repeatedly observed that small changes in prompt structure altered outcomes. For end-user contexts, it may be appropriate to evaluate not only accuracy on a fixed prompt set but also \emph{stability} under paraphrase, constraint restatement, and adversarial ambiguity.

\subsection{Implications for Model and Product Design}

\paragraph{Self-correction and calibration should be user-visible}
Students were especially attentive to moments when models contradicted themselves, defended an incorrect path, or appeared to ``argue'' rather than reassess. Many students reported that this behaviour genuinely surprised them, as they had never encountered it before.  Interfaces that surface uncertainty, invite verification, and support structured checking (rather than rhetorical elaboration) may improve real-world epistemic safety.

\paragraph{Constraint tracking is a practical priority}
Many failures documented by teams can be reframed as constraint-tracking failures: multi-step transformations, strict rule application, and spatial/visual reasoning with precise state updates.
Improvements in these areas would likely yield outsized benefits for end-user trustworthiness.

\subsection{Research Opportunities}

This course-embedded design suggests several research directions that are feasible without turning a classroom into a lab:

\begin{enumerate}[leftmargin=*,itemsep=0.25em]
\item \textbf{Multi-institution replication:} repeat the midterm protocol across universities and semesters to study how prompt creativity, student background, and model updates shape observed failure modes.
\item \textbf{Rubric standardization:} develop a shared scoring rubric with calibration prompts to improve inter-rater reliability and enable meta-analysis of results.
\item \textbf{Corpus release (where permitted):} publish an anonymized subset of prompts and responses as an ``end-user reasoning probes'' dataset, emphasizing interactional and explanation-level phenomena.
\item \textbf{Linking pedagogy to outcomes:} measure how participation changes students' later trust calibration and verification behaviors in authentic settings beyond the course.
\end{enumerate}

\section{Conclusion}
\label{sec:Conclusion}

This paper documents an unusual but increasingly necessary kind of work: a general-education course that teaches AI literacy by having students conduct disciplined evaluations of the AI systems they encounter in everyday life. The UNIV~182 midterm project operationalizes the contested question ``Can consumer chatbots reason?'' as a student-led field experiment with explicit protocols, a reasoning-task taxonomy, and a two-metric evaluation scheme separating answer correctness from explanation validity.

Three conclusions emerge.

\paragraph{First, the pedagogical conclusion: evaluation is teachable as practice}
Students across majors and prior STEM exposure successfully designed non-trivial reasoning probes, ran controlled comparisons across chatbots, logged outputs, computed summary statistics, and defended claims with evidence and caveats. In doing so, they demystified chatbots: systems that feel authoritative at first encounter became testable artifacts with identifiable strengths, weaknesses, and failure signatures.

\paragraph{Second, the empirical conclusion: performance is uneven across reasoning categories and fragile under constraint}
Across heterogeneous prompt sets, teams converged on a consistent pattern: high performance on short, structured quantitative and pattern tasks, and reduced reliability on spatial/visual reasoning, multi-step transformations, and prompts requiring strict constraint tracking. Just as importantly, students documented frequent mismatches between fluent explanations and valid justifications, motivating evaluation beyond surface-level correctness.

\paragraph{Third, the methodological conclusion: interactional behavior matters}
Consumer chatbots are not static question-answering systems; they are interactive products. Students observed that prompting, clarification, and challenge can elicit different behaviors, including overconfidence, persistence, and post-hoc rationalization.
These behaviors are central to real-world trust calibration and should be treated as part of ``reasoning'' evaluation in end-user settings.

Altogether, the midterm project demonstrates that ``AI-for-all'' education can be both technically meaningful and broadly accessible: students can learn core AI literacy skills by building and executing experiments, and the resulting artifacts can contribute a complementary perspective on how consumer chatbots behave in authentic use.

\section{Ethics, Consent, and Data Stewardship}
\label{sec:Ethics}

All student coauthors included on this manuscript provided explicit, affirmative consent to be listed as coauthors on a public preprint, collected by the course instructor. Students were informed of what public authorship implies (public visibility of names, long-term persistence of the manuscript online, and the possibility of downstream citation).

Because this work originates in a graded course, special care is required to avoid coercion. Consent to coauthorship was decoupled from grading and collected separately, with students informed that non-consent would not affect assessment. Only students with graded submissions were eligible for inclusion as coauthors, consistent with contribution-based authorship norms.

Chatbot outputs were collected from publicly available consumer interfaces. Platform terms of service, interface restrictions, and the evolving nature of model deployments constrain what can be redistributed. Accordingly, this paper prioritizes aggregate summaries and representative excerpts rather than full transcript release unless explicit permission and terms compliance are established.

Students were instructed to design reasoning tasks that are benign and educational, not to probe for unsafe capabilities, bypass safeguards, or elicit harmful content. The assignment goal is AI literacy through disciplined evaluation, not red-teaming for exploitation.

Finally, we treat reported performance numbers as descriptive summaries of heterogeneous student-designed evaluations, not as definitive claims about model superiority. This is an ethical stance as well as a methodological one: overclaiming from classroom data would risk misleading readers and misrepresenting what the evidence supports.

\section*{Acknowledgments}
The instructor of this course acknowledges and thanks all the students in UNIV 182, Fall 2025, for their sustained curiosity, professionalism, and enthusiasm during the semester.



\bibliographystyle{unsrtnat} 
\bibliography{references}

\newpage
\clearpage
\section*{Appendix}

\subsection*{Prompts}

Prompts below are aggregated across student written submissions and slide decks. Prompts are reproduced as used
by students, with minor normalization for readability and \LaTeX{} compatibility (e.g., rendering arrows as math
symbols). Where prompts originally used emojis, we retain the intended structure using short text descriptors in
brackets. Some teams provided a small number of prompts only as short descriptions (e.g., ``solve a cryptogram'');
these are preserved verbatim as phrased by the students.

\subsubsection*{Pattern Completion}

\begin{enumerate}[leftmargin=*]
  \item The sequence is $3, 6, 9, 12$: what number comes next, and why?
  \item The sequence is $2, 4, 8, 16$: what number comes next, and why?
  \item The sequence is $1, 2, 4, 7, 11, 16$. What comes next, and why?
  \item The letters go B, D, F, H. What is the next letter and why?
  \item What comes next in the pattern: A, C, F, J, O?
  \item The sequence of numbers is: $1, 11, 21, 1211, 111221, \ ?$ What is the next number in the sequence?
  \item The pattern of shapes is: square, circle, square, circle, square. What is the next shape?
  \item In a repeating symbol pattern $\rightarrow, \downarrow, \leftarrow, \uparrow$, what is the direction of the 9th arrow?
  \item You have the sequence of words: \emph{cat}, \emph{caterpillar}, \emph{category}. What would be the next word in the sequence and why? (Answer must have the \texttt{cat-} prefix.)
  \item Each word increases by one letter: \emph{go}, \emph{goo}, \emph{good}, \emph{goods}. What comes next if the rule continues?
  \item A number sequence proceeds as: $1, 8, 63$. What is the next number in the sequence?
  \item Each word in a list starts \emph{r}, \emph{ra}, \emph{rad}, \emph{radi}. What could the next two words be if this continues?
  \item Here is a pattern of color combinations: Yellow and Purple, Orange and Blue. What color combination should come next and why?
  \item The sequence is A2, C4, F16, and J256. What comes next, and what is the rule for both the letter and the number in this pattern?
  \item There is a pattern to the numbers and letters: A=1, B=28, C=55, D=82, E=109, F=136. If the next letter is G, what would its corresponding number be and what is your reasoning?
  \item If you say the full alphabet and skip each 2nd letter (a, c, e, \ldots), how many letters are there in the resulting sequence?
  \item In a repeating symbol pattern \texttt{+}, \texttt{*}, \texttt{=}, \texttt{+}, \texttt{*}, \texttt{=}, \ldots, what is the 11th symbol?
  \item M1N2 $\rightarrow$ M1N3, as AA11BB22 is to?
  \item In a repeating emoji pattern [laughing face], [upside-down face], [frowning face], [star-struck face], [star-struck face], \ldots, what would the 20th emoji be?
  \item The pattern of fruit is [broccoli], [grape], [orange], [orange], [banana], [watermelon], [strawberry], what should come next? 
  \item You have a pattern: Blue, green, 6, !, a, Weep, \ldots\ What comes next and why?
  \item Signal in noise: predict the next digit after a mostly random sequence with a hidden pattern. 
\end{enumerate}

\subsubsection*{Transformation Rules}

\begin{enumerate}[leftmargin=*]
  \item Every shape loses a side. A pentagon becomes a square and a square becomes a triangle. What does a septagon become after two transformations?
  \item Every triangle becomes a circle; every circle becomes a square (and, in one variant, every square becomes a triangle). What does a triangle become after two transformations?
  \item Replace each vowel in a word with the next vowel (a$\rightarrow$e, e$\rightarrow$i, i$\rightarrow$o, o$\rightarrow$u, u$\rightarrow$a). What does \emph{code} become?
  \item Replace every vowel in a word with the next vowel (a$\rightarrow$e, e$\rightarrow$i, i$\rightarrow$o, o$\rightarrow$u, u$\rightarrow$a). What does \emph{red} become?
  \item Transform the word \emph{smile}: for every vowel, insert the letter \texttt{p} before it; for every consonant, double it. What is the resulting word?
  \item Spell the sentence ``I ran with my friend Micheal and then we played basketttbal with my other friend Fernando'' backwards (character-level reversal; preserve misspellings).
  \item Each animal doubles its legs: spiders have $8 \rightarrow 16$ legs, cats $4 \rightarrow 8$. According to this rule, what happens to a bird?
  \item Each insect triples its legs: ants have $6 \rightarrow 18$ legs, spiders have $8 \rightarrow 24$. According to this rule, what happens to a fly?
  \item ABCD $\rightarrow$ ABCE. PPQQRRSS $\rightarrow$ ? 
  \item If an object is alone, it turns red; if next to another of the same color, it turns blue. Three red objects are touching. What color do they become?
  \item In a grid, all cells touching a black square turn black. Starting with one black square in the middle of a $3\times 3$ grid, how many black squares are black after one step?
  \item If ``ACE'' becomes ``BDF,'' how should ``CAT'' transform following the same rule?
  \item Solve a very short cryptogram that follows a double change. 
  \item Assign a word a score based on its sequences of consonants and vowels, then reverse and produce a word that will match a stated score. 
  \item A cross can be turned into a ribbon if bent, then back into a cross if the reverse happens. What would the result be if both were done at the same time?
  \item You can only move one matchstick to make this equation true: $6 + 4 = 4$. How do you fix it?
\end{enumerate}

\subsubsection*{Spatial \& Visual Reasoning}

\begin{enumerate}[leftmargin=*]
  \item Picture a $3\times 3$ grid with a dot in the top left. It moves one cell to the right each step. Where is it after four steps?
  \item Picture a $3\times 3$ grid with a dot in the top left. It moves one cell to the right each step. Where is the dot after six steps? 
  \item A row has 1 star in the first line, 2 in the next, 3 in the next, and so on. What will the fifth line look like?
  \item Imagine a staircase made of blocks: 3 on the bottom, 2 above those, 1 on top. If you flip it horizontally, what is its shape?
  \item A grid shows a black square mirror-reflected across the diagonal. Where does it move?
  \item A robot turns right every 90 degrees and moves 1 meter. After 4 moves, where is it relative to the start?
  \item Imagine a $5\times 5$ grid with a robot in the center. The robot moves forward one tile and turns right 90 degrees each turn. If it starts facing down, where will the robot be after 3 turns?
  \item A row has 2 trapezoids in the first line, 4 in the next, 6 in the next, and so on. What will the seventh line look like?
  \item Clock face at 3:00. Rotate the clock 45 degrees counterclockwise about the center. Then place a mirror along the 6--12 line. Describe the reflected positions.
  \item In a repeating emoji pattern (e.g., [circle], [square], [triangle], [circle], \ldots), what would the 20th symbol be?
  \item If walking is to galactic travel, what is warp speed to?  \item What is the best bridge to take to cross the river between MIT and Harvard?
  \item What’s the next country you hit heading due east from Nashville? 
  \item What’s the next country you hit heading due east from Knoxville? 
  \item Predict the next number in a series based on points on a sine curve.
  \item A cube has faces painted Red, Blue, Green, Yellow, White, and Black. The Red face is opposite the Green face. The Blue face is next to the Red face. The Blue face is opposite the Yellow face. If the White face is on the bottom, what color is the top face?
\end{enumerate}

\subsubsection*{Counting and Quantitative Reasoning}

\begin{enumerate}[leftmargin=*]
  \item If each green box counts as 1 point, and each blue box is 3 points, what is the total for 3 green and 4 blue boxes?
  \item If each blue tile counts 2 points and each yellow tile counts 3 points, what is the total for 4 blue and 2 yellow tiles?
  \item If each red tile counts 2 points and each yellow tile counts 3 points, what is the total for 5 red tiles and 3 yellow tiles?
  \item There are 5 shelves, each holding 10 books. Every third book on each shelf is hardcover. How many hardcover books are there?
  \item If there is \$100 in Richard's box, Harry steals \$34, Mary borrows \$40, and Bob takes \$55. What is the situation? (Be explicit about feasibility.)
  \item If there is \$100 in Richard's box, Harry steals \$34, Mary borrows \$40, and Bob takes \$55. What will Richard's reaction be? 
  \item What is $0.04694658 \times 4{,}662{,}437$?
  \item If each column triples in height from the previous column, and column 1 is 6 ft, what will the height be of the 4th column?
  \item If there is a pyramid of cubes with a base of 17, how many cubes will there be when the top is placed? 
  \item How many cubes have exactly two faces showing in a $3\times 3\times 3$ block? 
  \item There's a $3\times 3\times 3$ cube made up of $1\times 1\times 1$ cubes; if every side is painted, how many $1\times 1\times 1$ cubes are painted exactly twice? 
  \item You add consecutive odd numbers: $1$, $1+3$, $1+3+5$, $1+3+5+7$. What pattern emerges?
  \item If today is Monday and you add 100 days, what day of the week will it be?
  \item A farmer is building square corrals, counting only the perimeter of fence posts. A $1\times 1$ corral uses 4 posts. A $2\times 2$ corral uses 8 posts. A $3\times 3$ corral uses 12 posts. How many fence posts are needed for the tenth square corral in this pattern?
  \item Given the rule that every sea creature now has double the eyes that they had before (e.g., a great white shark now has 4 eyes), how many eyes does the Six-eyed Spookfish have?
  \item If a person has one hand and raises a finger every second, how many fingers do they have raised after 7 seconds? 
  \item If a student were to write the sentence ``We are incapable of locating the baggage carousel,'' how many letter ``a''s are there?
  \item A complex combination/permutation question. 
  \item Determine the correct number in a series with a simple but non-obvious pattern. 
\end{enumerate}

\subsubsection*{Relational \& Logic Reasoning}

\begin{enumerate}[leftmargin=*]
  \item All squares are heavier than circles, and all circles are heavier than triangles. Which shape is lightest?
  \item All pentagons are heavier than octagons, and octagons are heavier than trapezoids. Which shape is the lightest?
  \item If all green shapes are big and all big shapes are squares, what can you say about green shapes?
  \item Every student taller than Sam likes math. Alex likes math. Can you tell whether Alex is taller than Sam?
  \item No birds can swim. Penguins are birds. Can penguins swim? \item John is taller than Mary. Mary is taller than Kate. Kate is shorter than Bob, who is shorter than John. Who is the tallest?
  \item Bob is talking to Mary. Mary is talking to JoAnn. JoAnn is married to Bill. Who is Dan? 
  \item All cats are animals. Some animals are not pets. Are all cats pets? Explain your reasoning.
  \item Determine a method to find which object (of 12) has a different weight using a balance scale in only three weighings. \item Prove that you cannot prove this sentence. 
  \item Why can scientists predict when hurricanes will hit but not earthquakes?
  \item Humans can get dizzy from motion and fluid in their inner ear. The earth is constantly rotating, so why aren't humans dizzy living on planet earth?
  \item What are the most important differences between Germany in 1937 and the United States in 2025? 
  \item If sunlight causes cancer, does that mean darkness is objectively better? 
  \item If killing bugs is immoral, are picking flowers also immoral? Yes or no. 
  \item Imagine you remember every previous version of yourself. Which memory decides what ``you'' believe now: the first, the last, or the one that remembers remembering?
  \item After WWII, if Germany’s a magenta triangle, Japan a cerulean quadrilateral, and Russia a pink hexagon, what form and hue is America? 
  \item What was the best, most exciting NFL game last weekend? 
\end{enumerate}

\subsubsection*{Analogical Reasoning}

\begin{enumerate}[leftmargin=*]
  \item Student is to teacher, as child is to what?
  \item Cat is to mouse, as bird is to ?
  \item A cat is to a mouse as a bird is to a what? 
  \item Tree : leaf :: flower : ?
  \item Water is to beach as snow is to what?
  \item Chicken is to a bird as frog is to a what?
  \item Oar is to tree as quill is to ?
  \item As a key unlocks a door, a spark ignites a ?
  \item Hand : glove :: foot : ?
  \item In the context of mathematics, subtraction is to division what addition is to what?
  \item Hot is to cold as socks are to ?
  \item Broccoli is healthy; what does that make pizza?
  \item If ``flap'' means ``to fly'' and ``snap'' means ``fast,'' what does \emph{flap snap} likely mean?
  \item Create an analogy to a pair of seemingly unrelated words. 
  \item 3 dimensions is to a triangle as 15 dimensions is to what? 
\end{enumerate}

\end{document}